\documentclass[superscriptaddress,12pt,showpacs,pra]{revtex4}

\usepackage[tbtags]{amsmath}
\usepackage{threeparttable}
\usepackage{amssymb}
\usepackage{amsfonts}
\usepackage{graphicx}
\usepackage{xcolor}
\usepackage[dvipdfmx]{hyperref}
\hypersetup{colorlinks=true,breaklinks=true,pdfstartview=FitH,linkcolor=blue}
\usepackage{tabularx}

\begin{document}

\title{Non-homogeneous Bell-type Inequalities for Two- and Three-qubit States}
\author{Mingjun Shi}%
\affiliation{Department of Modern Physics, University of Science and Technology of China, Hefei,
Anhui 230026, People's Republic of China}
\author{Changliang Ren}
\affiliation{Department of Modern Physics, University of Science and Technology of China, Hefei,
Anhui 230026, People's Republic of China}
\author{Jiangfeng Du}
\email{djf@ustc.edu.cn}%
\affiliation{Department of Modern Physics, University of Science and
Technology of China, Hefei, Anhui 230026, People's Republic of China} \affiliation{Hefei National
Laboratory for Physical Sciences at Microscale, Hefei, Anhui 230026, People's Republic of China}

\begin{abstract}
A systematic approach is presented to construct non-homogeneous two- and three-qubit Bell-type
inequalities. When projector-like terms are subtracted from homogeneous two-qubit CHSH polynomial,
non-homogeneous inequalities are attained and the maximal quantum mechanical violation
asymptotically equals a constant with the subtracted terms becoming sufficiently large. In the case
of three-qubit system, it is found that most significant three-qubit inequalities presented in
literature can be recovered in our framework. We aslo discuss the behavior of such inequalities in
the loophole-free Bell test and obtain corresponding thresholds of detection efficiency.
\end{abstract}

\maketitle

%================================================================================

\section{Introduction}
Bell's original inequality reveals the conflict of quantum mechanics and local hidden variable
(LHV) theory \cite{Bell 1964}. Since then, various forms of Bell-type inequalities have been
derived. The most well-known is Clauser-Horne-Shimony-Holt (CHSH) inequality \cite{CHSH}.
Mermin-Ardehali-Belinskii-Klyshko (MABK) generalized the result to the case of $N$ qubits
\cite{MABK}. Subsequently, Werner and \.{Z}ukowski (WW\.{Z}B) presented explicit description of
$N$-qubit Bell-correlation inequalities for two dichotomic observables per site \cite{Werner
2001,Zukowski 2002-1}. It should be stressed that all these results only concern with the ``full''
correlation function, that is, the expectation value of the product of all $N$ local observables.
We call this class of Bell-type inequalities as \emph{homogeneous} inequalities.

Besides full correlations, one can take into account partial correlations, which involve the
product of not $N$ but $n<N$ observables. It is meaningful to construct such \emph{non-homogeneous}
Bell-type inequalities that incorporate partial correlations as well as full correlations. The main
reason is as follows. Gisin proved that the bipartite pure entangled states violate CHSH inequality
\cite{Gisin,Gisin and Peres}. The problem whether Gisin theorem can be generalized for an arbitrary
n-partite pure entangled states remains open. In the three-qubit case there are generalized GHZ
states that do not violate the MABK inequalities \cite{Scarani,Zukowski 2002-2}. More generally it
has been shown that these states do not violate any Bell-type inequality for n-partite correlation
functions (that is, homogeneous inequality) for experiments involving two dichotomic observables
per site \cite{Zukowski 2002-2}. Subsequently some Bell-type inequalities, which bear the
non-homogeneous form, were presented and can be violated by generalized GHZ states \cite{J
Chen,Chen-Kai}. Besides, non-homogeneous inequalities have appeared in the discussion of the
non-locality of cluster states \cite{cluster, Scarani2005}. However one can hardly make clear how
and why these inequalities come out. In other words, there is no systematic formulation of such
inequalities.

Motivated by this issue, we present in this paper a feasible method to construct non-homogeneous
Bell-type inequalities. The key point is so simple that we can describe it very briefly in one
sentence, that is, non-homogeneous inequalities are derived by subtracting some projector-like
terms from the homogeneous ones.

We will first present the construction of two-qubit non-homogeneous inequalities. The newly
obtained inequalities are weaker than CHSH inequality in that the quantum mechanical violation of
the former is smaller than that of the latter. Nevertheless, they pave the way for the considering
more general cases. Consequently we attain so many three-qubit non-homogeneous inequalities and
some meaningful results. In Ref. \cite{Pitowsky}, Pitowsky and Svozil have presented several
optimal three-qubit Bell-type inequalities in the sense that they represent the best possible upper
bounds for the conceivable classical probabilities. These inequalities are filtrated from tens of
thousands of inequalities describing the facets of classical correlation polytope. Though these
inequalities appear very intricate, they can be constructed systematically in our framework.
Additionally the two Bell-type inequalities, which were presented in Ref. \cite{J Chen} to disclose
the non-locality of generalized three-qubit GHZ states, are included in our results.

When these inequalities are utilized to display the non-locality experimentally, one has to
consider the effect of non-ideal detector. Using the software Mathematica and MatLab, we
numerically analyze the detection efficiency limit of the above-mentioned three-qubit Bell-type
inequalities. Three detectors may have the same efficiency or not \cite{Cabello and Brunner}. With
respect to these two different cases, we attain the thresholds of detection efficiency. Some
inequalities have the advantage that the efficiency of one detector can be arbitrarily low if that
of the other two detectors satisfy certain conditions.

\section{Two-qubit Non-homogeneous Bell-type Inequalities}

We start by briefly recalling LHV model and Bell-type inequalities. Bell-type inequalities always
refer to correlations between two or more sites. In the two-qubit case, each of two space-separated
observers, A and B, gets a qubit and measures two $\pm1$-valued local observables, denoted $A_1$,
$A_2$ for observer A, and $B_1$, $B_2$ for observer B. The outcomes of measurement are labeled by
$\textsf{a}_i=\pm1$ and $\textsf{b}_j=\pm1$ for $i,j=1,2$.

A typical joint probability can be expressed as $P(\textsf{a}_2,\textsf{b}_1|A_2,B_1)$, where after
the vertical bar we write the observables chosen at two sites, and before the bar the particular
outcomes.

In the formalism of local hidden variable (LHV) theory, there exists a hidden variable $\lambda$
which takes values in space $\Lambda$. With the presence of $\lambda$, the probability of measuring
$A_i$ and obtaining the outcome $\textsf{a}_i$ is represented by $P(\textsf{a}_i|A_i,\lambda)$.
Similarly for $P(\textsf{b}_j|B_j,\lambda)$. Given $\lambda$, one can calculate the mean value of
$\textsf{a}_i(\lambda)$ from the probability $P(\textsf{a}_i|A_i,\lambda)$,
\begin{equation}\label{a bar lamda}
\bar{\textsf{a}}_i(\lambda)=P(+1|A_i,\lambda)-P(-1|A_i,\lambda).
\end{equation}
Similarly for $\bar{\textsf{b}}_i(\lambda)$.

Furthermore LHV theory requires that with the presence of $\lambda$ the joint probability
$P(\textsf{a}_i,\textsf{b}_j|A_i,B_j,\lambda)$ is factorisable, that is,
\begin{equation}\label{Bell condition}
P(\textsf{a}_i,\textsf{b}_j|A_i,B_j,\lambda)=P(\textsf{a}_i|A_i,\lambda)P(\textsf{b}_j|B_j,\lambda),
\end{equation}
where $P(\textsf{a}_i|A_i,\lambda)$ is independent of observable $B_j$ and its outcome
$\textsf{b}_j$, and $P(\textsf{b}_j|B_j,\lambda)$ independent of $A_i$ and $\textsf{a}_i$.

Given the probability measure $\mu$ on $\Lambda$, using \eqref{Bell condition}, one can compute the
joint probability
\begin{equation}
P(\textsf{a}_i,\textsf{b}_j|A_i,B_j)=\int_{\Lambda}P(\textsf{a}_i|A_i,\lambda)P(\textsf{b}_j|B_j,\lambda)\mu(\lambda)\,d\lambda.
\end{equation}
Consequently the expectation value of joint measurement is
\begin{equation}
\begin{split}
&\langle A_iB_j\rangle_{\mathrm{lhv}}
 =\sum_{\textsf{a}_i,\textsf{b}_j}\textsf{a}_i\textsf{b}_j\,P(\textsf{a}_i,\textsf{b}_j|A_i,B_j)  \\
=&\int_{\Lambda}\sum_{\textsf{a}_i,\textsf{b}_j}\textsf{a}_i\textsf{b}_j\,P(\textsf{a}_i|A_i,\lambda)
                   P(\textsf{b}_j|B_j,\lambda)\mu(\lambda)\,d\lambda \\
=&\int_{\Lambda}\bar{\textsf{a}}_i(\lambda)\bar{\textsf{b}}_j(\lambda)\mu(\lambda)\,d\lambda.
\end{split}
\end{equation}

In the context of hidden variable model, correlation function is defined as
\begin{equation}\label{E lambda}
\begin{split}
E(\lambda)=\bar{\textsf{a}}_1(\lambda) & \bar{\textsf{b}}_1(\lambda)-\bar{\textsf{a}}_1(\lambda)\bar{\textsf{b}}_2(\lambda)\\
              &  -\bar{\textsf{a}}_2(\lambda)\bar{\textsf{b}}_1(\lambda)-\bar{\textsf{a}}_2(\lambda)\bar{\textsf{b}}_2(\lambda).
\end{split}
\end{equation}
One can see that $|E(\lambda)|\leq2$, and the inequality is saturated only if
$\bar{\textsf{a}}_i(\lambda)$ and $\bar{\textsf{b}}_j(\lambda)$ take the extremal values of $\pm1$.
CHSH inequality is attained by averaging over hidden variable $\lambda$.
\begin{equation}
\big|\langle A_1B_1\rangle_{\mathrm{lhv}}-\langle A_1B_2\rangle_{\mathrm{lhv}}-\langle
A_2B_1\rangle_{\mathrm{lhv}}-\langle A_2B_2\rangle_{\mathrm{lhv}}\big|\leq 2.
\end{equation}

In the following discussion, it is convenient to adopt $|E(\lambda)|\leq 2$ to represent CHSH
inequality, that is,
\begin{equation}\label{CHSH inequality}
    -2\leq a_1b_1-a_1b_2-a_2b_1-a_2b_2\leq 2,
\end{equation}
where we replace $\bar{\textsf{a}}_i(\lambda)$ and $\bar{\textsf{b}}_j(\lambda)$ with $a_i$ and
$b_j$ for simplicity. We also make the following definitions.
\begin{align}
 & E_1=a_1b_1-a_1b_2-a_2b_1-a_2b_2,\\
 & E_2=-a_1b_1+a_1b_2-a_2b_1-a_2b_2,\\
 & E_3=-a_1b_1-a_1b_2+a_2b_1-a_2b_2,\\
 & E_4=-a_1b_1-a_1b_2-a_2b_1+a_2b_2.
\end{align}
Each $E_k$ is called (homogeneous) CHSH polynomial and has the same bound of $2$, i.e., $|E_k|\leq
2$ for $k=1,2,3,4$.

\subsection{Method}

Now we start to construct non-homogeneous polynomial, denoted $E'$. We require that $E'$ satisfies
two conditions:
\begin{itemize}
  \item[(i)] $E'\leq 2$. That means in the LHV upper bound of $E'$ is $2$.
  \item[(ii)] In the case of quantum mechanics, $E'$ is replaced with the operator form
  $\mathcal{E}'$, and the expectation value given by $\langle\mathcal{E}'\rangle_{\mathrm{qm}}$ can
  be larger than 2.
\end{itemize}
In other words, $E'\leq 2$ is a Bell-type inequality.

The key point of the method presented here is so transparent that at first sight it seems to be
trivial: since every CHSH polynomial satisfies $E_k\leq 2$, it is evident that $E_k$ minus some
non-negative terms must be equal to or less than $2$. These non-negative terms are chosen to be
\begin{equation} \label{LHV Pa and Pb}
P_{a_{i}}^{\pm}=\frac{1\pm a_{i}}{2},\quad P_{b_{j}}^{\pm}=\frac{1\pm b_{j}}{2}.
\end{equation}
A typical non-homogeneous polynomial is then given by
\begin{equation} \label{E prime}
E'(r)=E_{1}-rP_{a_{2}}^{+}P_{b_{2}}^{+},
\end{equation}
where $r$ is any non-negative number.

Considering condition (i), we note that $E'\leq 2$ and $E'$ reaches the maximal value of $2$ at
specific extremal values of $a_i$ and $b_j$. For example, when $a_1=1$, $a_2=1$, $b_1=1$ and
$b_2=-1$, $E'$ equals $2$. So \eqref{E prime} satisfies condition (i).

We now show that condition (ii) can also be fulfilled. In quantum-mechanical case, $a_i$ and $b_j$
in \eqref{LHV Pa and Pb} are replaced with spin observables, namely, $a_i\Rightarrow A_i$,
$b_j\Rightarrow B_j$. Correspondingly, we have
\begin{gather}
    P_{a_i}^{\pm}\Rightarrow\mathcal{P}_{a_i}^{\pm}=\tfrac{I\pm A_{i}}{2},
    \quad P_{b_j}^{\pm}\Rightarrow\mathcal{P}_{b_{j}}^{\pm}=\tfrac{I\pm B_{j}}{2};\\
    E_1\Rightarrow \mathcal{E}_1=A_1B_1-A_1B_2-A_2B_1-A_2B_2,
\end{gather}
where $I$ is $2\times2$ identity matrix.

Then the quantum mechanical form of \eqref{E prime} is
\begin{equation} \label{quantum E prime}
\mathcal{E}'(r)=\mathcal{E}_{1}-r\mathcal{P}_{a_{2}}^{+}\mathcal{P}_{b_{2}}^{+},
\end{equation}
Any two-qubit entangled pure state can be expressed as the form of Schmidt decomposition (see, for
example, \cite{Nielsen}),
\begin{equation}\label{Psi with epsilon}
|\Psi\rangle=\cos\xi|00\rangle+\sin\xi|11\rangle, \quad \xi\in(0,\pi\texttt{/}2).
\end{equation}
We let
\begin{equation}\label{observables type 1}
\left.
\begin{array}{l}
A_{1}=\sigma_{x}, \quad A_{2}=-\sigma_{z}, \\
B_{1}=\sigma_{x}\sin\theta+\sigma_{z}\cos\theta,  \\
B_{2}=-\sigma_{x}\sin\theta+\sigma_{z}\cos\theta.
\end{array}
\right\}
\end{equation}
The expectation value of $\mathcal{E}'$ is given by
\begin{equation}
\begin{split}
\langle\Psi|\mathcal{E}'|\Psi\rangle & =\left(2+\frac{r}{2}\sin^{2}\xi\right)\cos\theta \\
 & +2\sin2\xi\sin\theta-\frac{r}{2}\sin^{2}\xi.
\end{split}
\end{equation}
When $\theta=\arctan\frac{4\sin2\xi}{4+r\sin^{2}\xi}$, $\langle\Psi|\mathcal{E}'|\Psi\rangle$
acquires the maximal value
\begin{equation}
\langle\Psi|\mathcal{E}'|\Psi\rangle_{\max}=\left[\left(2+\frac{r}{2}\sin^{2}\xi\right)^2
+4\sin^22\xi\right]^{1/2}-\frac{r}{2}\sin^{2}\xi,
\end{equation}
which is obviously larger than $2$. Therefore \eqref{E prime} represents a series of
non-homogeneous inequalities and can be used to detect non-locality.

The above procedure can be formulated in more general forms. Let's consider the following
non-homogeneous polynomial.
\begin{equation} \label{E 2prime}
E''(s,t)=E_{4}-sP_{a_{1}}^{+}P_{b_{2}}^{+} -tP_{a_{2}}^{+}P_{b_{1}}^{+},
\end{equation}
where $s$ and $t$ are non-negative numbers. This time we select $E_{4}$ to construct $E''$. It is
just for the later convenience and not necessary. Obviously we have $E''\leq 2$ and the bound is
attained for some extremal points, say, $a_1=a_2=1$ and $b_1=b_2=-1$. In order to show the
violation of $E''\leq 2$ in quantum mechanics, we evaluate the expectation value
$\langle\Psi|\mathcal{E}''|\Psi\rangle$, where $|\Psi\rangle$ is given by \eqref{Psi with epsilon}
and $\mathcal{E}''$ is the quantum  conterpart of $E''$.

For $\xi\in[0,\pi\texttt{/}4]$, we choose the observables as
\begin{equation}\label{observables type 2}
\left.
\begin{array}{l}
A_{1}=\sigma_{z}, \quad A_{2}=-\sigma_{x}, \\
B_{1}=\sigma_{x}\sin\theta+\sigma_{z}\cos\theta,  \\
B_{2}=-\sigma_{x}\sin\theta+\sigma_{z}\cos\theta.
\end{array}
\right\}
\end{equation}
For this choice, we have
\begin{equation}
\begin{split}
 & \langle\Psi|\mathcal{E}''|\Psi\rangle \\
=& -\frac{1}{4}\big[s+8+(s+t)\cos2\xi\big]\cos\theta \\
 & \quad +\frac{t+8}{4}\sin2\xi\sin\theta
    -\frac{1}{4}(s+t+s\cos2\xi)
\end{split}
\end{equation}
The maximum of $\langle\Psi|\mathcal{E}''|\Psi\rangle$ over all $\theta$ is given by
\begin{equation}
\begin{split}
& f_{1}(s,t,\xi)=\max\limits_{\mathrm{all}\
\theta}\langle\Psi|\mathcal{E}''|\Psi\rangle  \\
= & -\frac{1}{4}(s+t+s\cos2\xi)  \\
  & +\frac{1}{4}\big[\left(8+s+(s+t) \cos2\xi\right)^{2}+(8+t)^2
\sin^{2}2\xi\big]^{1/2}.
\end{split}
\end{equation}
It can be found that if $s$ and $t$ satisfy
\begin{equation}\label{st condition 1}
32-st+(32-st+8t)\cos2\xi >0, \quad \xi\in[0,\pi\texttt{/}4],
\end{equation}
the function $f_1(s,t,\xi)$ will be larger than $2$ (when $\xi=0$, $f_1=2$).

For $\xi\in[\pi\texttt{/}4,\pi\texttt{/}2]$, we choose
\begin{equation}\label{observables type 3}
\left.
\begin{array}{l}
 A_{1}=-\sigma_{z}, \quad A_{2}=-\sigma_{x}, \\
 B_{1}=\sigma_{x}\sin\theta+\sigma_{z}\cos\theta,  \\
 B_{2}=-\sigma_{x}\sin\theta+\sigma_{z}\cos\theta.
\end{array}
\right\}
\end{equation}
Similarly, we have
\begin{equation}
\begin{split}
& f_{2}(s,t,\xi)=\max\limits_{\mathrm{all}\
\theta}\langle\Psi|\mathcal{E}''|\Psi\rangle  \\
= & -\frac{1}{4}(s+t-s\cos2\xi)  \\
&  +\frac{1}{4}\left[\left(8+s-(s+t)\cos2\xi\right)^{2}+(8+t)^2 \sin^{2}2\xi\right]^{1/2}.
\end{split}
\end{equation}
If $s$ and $t$ satisfy
\begin{equation}\label{st condition 2}
32-st-(32-st+8t)\cos2\xi >0, \quad \xi\in[\pi\texttt{/}4,\pi\texttt{/}2],
\end{equation}
then $f_2(s,t,\xi)$ is larger than $2$ (when $\theta=\pi/2$, $f_2=2$).

It can be seen from \eqref{st condition 1} and \eqref{st condition 2} that both of $f_{1}$ and
$f_{2}$ are larger than $2$ if $st<32$. So, under this condition, $\langle\mathcal{E}''\rangle$
violate LHV upper bound.

Some remarks are needed here. In the above calculation, we do not intend to find the maximal
violation of LHV bound. The choices of observable, given by \eqref{observables type 1},
\eqref{observables type 2} and \eqref{observables type 3}, do not cover all possible local
measurements. They are so selected as to simplify the calculation. Therefore
$\langle\Psi|\mathcal{E}'|\Psi\rangle_{\mathrm{max}}$ and
$\langle\Psi|\mathcal{E}''|\Psi\rangle_{\mathrm{max}}$ do not represent the maximal violation of
LHV bound. In the same way, the condition of $st<32$ is just required in the case considered. In
the following subsection, we discuss the maximal violation of inequalities $E'\leq 2$ and $E''\leq
2$.

\subsection{Violations of inequality}
Two typical forms of non-homogeneous inequality have been presented in \eqref{E prime} and \eqref{E
2prime}. We now analyze the violation of them in quantum mechanics. The quantum bound can be
computed by means of Min-Max principle \cite{Halmos}, which states that for self-adjoin
transformations the operator norm is bounded by the minimal and maximal eigenvalues. For the
purpose of clarity, we choose the directions of local measurement to be orthogonal. In other words,
$A_1$ and $A_2$ anti-commute. So do $B_1$ and $B_2$. Furthermore, we let
\begin{equation}
A_1=B_1=\sigma_x, \;\;A_2=B_2=\sigma_y.
\end{equation}
The operator $\mathcal{E}'(r)$, which is quantum correspondent of $E'(r)$, is written as
\begin{equation}
\mathcal{E}'(r)=
\begin{pmatrix}
  -1-r & -1 & -1 & 1 \\
   -1  &  1 &  1 & 1 \\
   -1  &  1 &  1 & 1 \\
    1  &  1 &  1 & -1
\end{pmatrix}.
\end{equation}
The eigenvalue of $\mathcal{E}'(r)$, denoted $\varepsilon(r)$, is determined by the equation
\begin{equation}\label{eign equation of E prime}
\varepsilon^4+r\varepsilon^3-(r+8)\varepsilon^2-4r\varepsilon=0.
\end{equation}
Among the solutions of \eqref{eign equation of E prime}, the largest one must be between $2$ and
$3$. To see this, note that the left-hand side of \eqref{eign equation of E prime} is negative when
$\varepsilon=2$, while positive and monotone increasing when $\varepsilon\geq 3$.

\begin{figure}[tbph]
\begin{center}
\includegraphics[width=0.8\columnwidth]{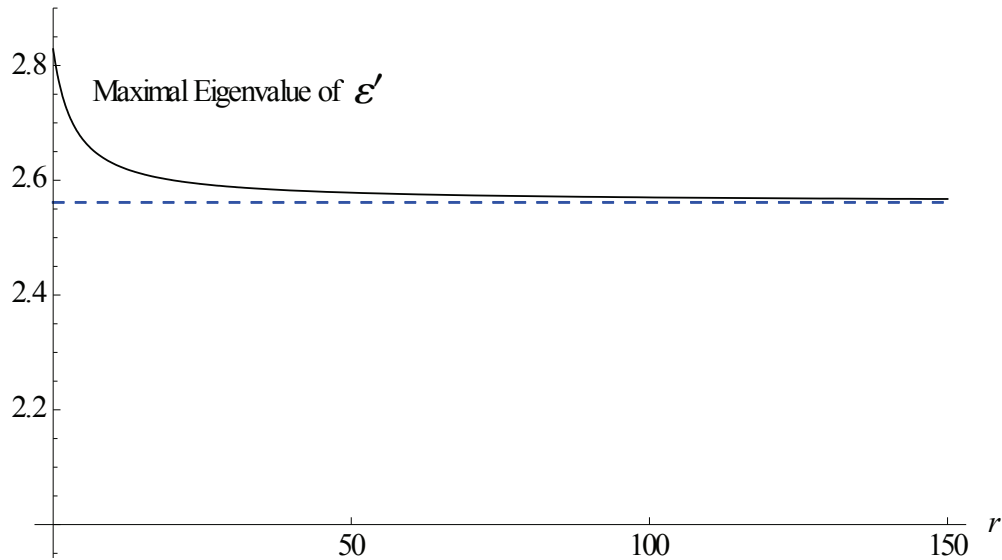}
\end{center}
\caption{(Color online) The maximal eigenvalue of $\mathcal {E}'(r)$. When $r$ tends to infinity,
the asymptotic value is $\frac{1+\sqrt{17}}{2}\approx2.56$.} \label{fig 1}
\end{figure}

The numerical result of the largest eigenvalue of $\mathcal{E}'(r)$ is plotted in Fig. \ref{fig 1}.
It can be seen that the largest eigenvalue asymptotically equals a constant (about $2.5$) for
sufficiently large $r$. The asymptotic value can be calculated from \eqref{eign equation of E
prime}. In fact, when $r\to \infty$, the terms containing $r$ should be finite. It follows that
\begin{equation}
\varepsilon^3-\varepsilon^2-4\varepsilon=0.
\end{equation}
The solution $\frac{1+\sqrt{17}}{2}\approx 2.56$ is what we want.

Similar analysis shows that the largest eigenvalue of $\mathcal{E}''(s,t)$ is also between $2$ and
$3$, and asymptotically equals $2.43$ when $s,t\to\infty$.

In a word, the inequalities $E'(r)\leq 2$ and $E''(s,t)\leq 2$ can be violated by quantum mechanics
for arbitrary non-negative numbers of $(r,s,t)$. The maximal violation corresponds to the case of
$r=s=t=0$. With $r$, $s$ and $t$ sufficiently large, the violation is close to a constant.

%-----------------------------------------------------------------------------------

\section{Three-qubit Non-Homogeneous Inequalities}
Based on the results obtained previously, we will in this section construct three-qubit
non-homogeneous inequalities.
\subsection{General results}
We now introduce the third qubit C. Two observables $C_{1}$ and $C_{2}$ are also $\pm1$-valued.
Just like Eq. \eqref{a bar lamda}, we define
\begin{equation}
c_i=\bar{\textsf{c}}_i(\lambda)=P(+1|C_i,\lambda)-P(-1|C_i,\lambda).
\end{equation}

The general form of three-qubit non-homogeneous polynomial can be expressed as
\begin{equation} \label{general form of 3-qubit}
F(a,b)c_{1}+G(a,b)c_{2}+H(a,b),
\end{equation}
where $F,\,G$ and $H$ are polynomials associated with qubit A and B. We will construct three-qubit
inequalities by the following steps. Firstly, we set the upper bound of \eqref{general form of
3-qubit} to be an arbitrary number, for example,
\begin{equation}\label{less than 2+u}
F(a,b)c_{1}+G(a,b)c_{2}+H(a,b)\leq 2+u,
\end{equation}
where we let $u\geq 0$ simply for the technical simplicity. Secondly we let $c_1$ and $c_2$ take
extremal values of $\pm 1$. \eqref{less than 2+u} will reduce to four inequalities with the upper
bound of $2+u$. We let three of them be the known two-qubit Bell-type inequalities (homogeneous or
non-homogeneous). For example, we make the following choice.
\begin{align}
& F+G+H=E''+u\leq 2+u, \label{reduced 1} \\
& F-G+H=E'+u\leq 2+u,  \label{reduced 2}\\
& -F-G+H=\tfrac{2+u}{2}(-E_4)\leq 2+u. \label{reduced 3}
\end{align}
So $F$, $G$ and $H$ can be expressed in terms of $E'(r)$, $E''(s,t)$ and $E_4$.
\begin{align}
& F=\frac{2+u}{4}E_4+\frac{E'}{2}+\frac{u}{2}, \label{F} \\[5pt]
& G=\frac{E''-E'}{2}, \label{G} \\[5pt]
& H=-\frac{2+u}{4}E_4+\frac{E''}{2}+\frac{u}{2}. \label{H}
\end{align}
Finally, there remains a reduced inequality, namely, $-F+G+H\leq 2+u$. We use it to determine the
possible values of $r$, $s$ and $t$. From \eqref{F}, \eqref{G} and \eqref{H}, it follows that
\begin{equation}\label{-F+G+H}
-F+G+H=-\frac{2+u}{2}E_4+E''-E'.
\end{equation}
The requirement that the upper bound of \eqref{-F+G+H} is $2+u$ leads to the following restrictions
on $r$, $s$ and $t$.
\begin{subequations}\label{conditions for rst}
\begin{equation}
r\leq 4+2u, \label{condition 1 for rst}
\end{equation}
\begin{equation}
r-s\leq 2u, \label{condition 2 for rst}
\end{equation}
\begin{equation}
r-t\leq 2u, \label{condition 3 for rst}
\end{equation}
\begin{equation}
r-s-t\leq 0. \label{condition 4 for rst}
\end{equation}
\end{subequations}
In a word, we require that four reduced two-qubit polynomials, namely $\pm F\pm G+H$, have the same
upper bound of $2+u$. We point out that the choice given by \eqref{reduced 1}--\eqref{reduced 3} is
not unique. What is more, some reduced inequality are not necessarily required to be Bell-type
inequalities, as will be seen in the next subsection.

\subsection{Specific inequalities}

We present some examples to illustrate the application of our method.
\subsubsection{$u=0$, $r=s=t=0$}
It follows that
\begin{gather}
E'=E_1,\;\; E''=E_4, \\
F=\frac{1}{2}(E_4+E_1),\;G=\frac{1}{2}(E_4-E_1),\;H=0.
\end{gather}
Then we have
\begin{equation}
-a_1b_1c_2-a_1b_2c_1-a_2b_1c_1+a_2b_2c_2\leq 2.
\end{equation}
It is the well-known three-qubit MABK inequality. The largest eigenvalue of the quantum
correspondent is $4$, which means the violation factor is $\frac{4}{2}=2$.

\subsubsection{$u=2$, $r=8,\,s=t=4$}

In this case, conditions \eqref{conditions for rst} are satisfied and saturated. Direct calculation
gives
\begin{equation} \label{Pito5}
\begin{split}
& -a_{1}-a_{2}-b_{1}-b_{2}+a_{1}b_{1}-a_{2}b_{2}-a_{1}c_{2}\\
& -2a_{2}c_{1}+a_{2}c_{2}-b_{1}c_{2}-2b_{2}c_{1}+b_{2}c_{2} \\
& -a_{1}b_{1}c_{1}-2a_{1}b_{1}c_{2}-3a_{1}b_{2}c_{1}-a_{1}b_{2}c_{2}\\
& -3a_{2}b_{1}c_{1}-a_{2}b_{1}c_{2}-a_{2}b_{2}c_{1}+4a_{2}b_{2}c_{2}\;\leq\, 8.
\end{split}
\end{equation}

In Ref. \cite{Pitowsky}, Pitowsky and Svozil enumerate some new Bell-type inequalities. In their
notation, one of the inequalities (Eq. (5) in \cite{Pitowsky}) is
\begin{equation} \label{Pito5 prob-form}
\begin{split}
& -P(A_{1})-2P(B_{1})-2P(C_{1})+2P(A_{1},B_{1})\\
& +2P(A_{1},C_{1})+P(A_{1},B_{2})+P(A_{1},C_{2})+P(A_{2},B_{1})\\
& +P(A_{2},C_{1})-P(A_{2},B_{2})-P(A_{2},C_{2})+2P(B_{1},C_{1})\\
& +2P(B_{2},C_{1})+2P(B_{1},C_{2})-2P(B_{2},C_{2}) \\
& -P(A_{1},B_{1},C_{1})-2P(A_{2},B_{1},C_{1})-3P(A_{1},B_{2},C_{1})\\
& -3P(A_{1},B_{1},C_{2})-P(A_{2},B_{2},C_{1})-P(A_{2},B_{1},C_{2}) \\
& -P(A_{1},B_{2},C_{2})+4P(A_{2},B_{2},C_{2})\quad\leq0.
\end{split}
\end{equation}
Here $P(A_1)$ denotes the probability that observer A measures observable $A_1$ and obtains the
outcome $+1$, that is, in our notation, $P(\textsf{a}_1=+1|A_1)$. We know that
\begin{equation}
\begin{split}
P(\textsf{a}_1=+1|A_1) & =\int_{\Lambda}P(\textsf{a}_1=+1|A_1,\lambda)\mu(\lambda)\,d\lambda\\[5pt]
  & =\int_{\Lambda}\frac{1+\bar{\textsf{a}}_1(\lambda)}{2}\mu(\lambda)\,d\lambda\\[5pt]
  & = \int_{\Lambda}\frac{1+a_1}{2}\mu(\lambda)\,d\lambda.
\end{split}
\end{equation}
Similarly $P(A_1,B_1)$ refers to $P(\textsf{a}_1=+1,\textsf{b}_1=+1|A_1,B_1)$. And we have
\begin{equation}
\begin{split}
 & P(\textsf{a}_1=+1,\textsf{b}_1=+1|A_1,B_1) \\[5pt]
=& \int_{\Lambda}
P(\textsf{a}_1=+1,\textsf{b}_1=+1|A_1,B_1,\lambda)\mu(\lambda)\,d\lambda\\[5pt]
=& \int_{\Lambda}P(\textsf{a}_1=+1|A_1,\lambda)P(\textsf{b}_1=+1|B_1,\lambda)\mu(\lambda)\,d\lambda\\[5pt]
=& \int_{\Lambda}\frac{1+a_1}{2}\frac{1+b_1}{2}\mu(\lambda)\,d\lambda.
\end{split}
\end{equation}
Then the inequality \eqref{Pito5 prob-form} can be rewritten in terms of $a_i,b_j,c_k$.

We find that, under permutation $b\rightarrow a$, $a\rightarrow c$ and $c\rightarrow b$, Pitowsky's
inequality \eqref{Pito5 prob-form} can be transformed to \eqref{Pito5}. Furthermore, the other
inequalities presented in \cite{Pitowsky} can be recovered in our framework.

In quantum mechanics, the largest eigenvalue of the operator corresponding to the left-hand side of
\eqref{Pito5} is about $12.87$ (numerically), which means the violation factor is about
$\frac{12.87}{8}\approx1.61$.

\begin{figure}[tbph]
\begin{center}
\includegraphics[width=0.8\columnwidth]{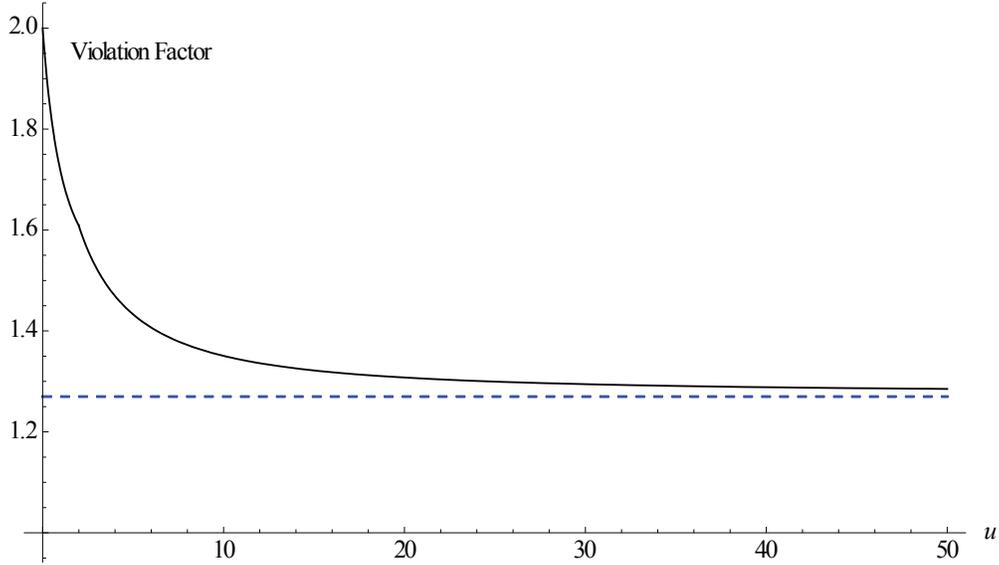}
\end{center}
\caption{(Color online) The maximal violation factor for different $u$. When $u$ sufficiently
large, the asymptotic value is about $1.27$.} \label{fig 2}
\end{figure}

We also evaluate various violation factors for different values of $(u,r,s,t)$. Numerical results
reveal such phenomena: (i) When $u\leq 2$, the optimal inequality, which can give rise to the
maximal violation of LHV bound, corresponds to the case of $r=4u$ and $s=t=2u$, that is, conditions
\eqref{condition 2 for rst}, \eqref{condition 3 for rst} and \eqref{condition 4 for rst} are
saturated. (ii) when $u\geq 2$, the optimal inequality is attained for $r=4+2u$ and $s=t=2+u$, that
is, conditions \eqref{condition 1 for rst} and \eqref{condition 4 for rst} are saturated. In Fig.
\ref{fig 2}, we plot the maximal violation factor for different values of $u$. It can be seen that
the violation factor tends to be a constant (about $1.27$) for large $u$.

\subsubsection{Another case}

Let's recall the steps by which the three-qubit non-homogeneous inequalities are established. The
key point is to let three reduced polynomials be related to three two-qubit Bell-type inequalities,
which is demonstrated in Eqns. \eqref{reduced 1}, \eqref{reduced 2} and \eqref{reduced 3}. This
requirement can be relaxed. For example, we consider the following procedure.
\begin{align}
& F+G+H=2-P_{a_1}^+P_{b_2}^+-P_{a_2}^+P_{b_1}^+\leq 2, \label{chen2 eqn 1} \\
& -F+G+H=E_4\leq 2, \label{chen2 eqn 2}\\
& -F-G+H=2-P_{a_1}^-P_{b_1}^--P_{a_2}^-P_{b_2}^-\leq 2. \label{chen2 eqn 3}
\end{align}
The LHV upper bounds of above polynomials are the same. But only \eqref{chen2 eqn 2} is a real
Bell-type inequality. From these equations, we get the following inequality.
\begin{equation}
\begin{split}
& -a_1b_1-a_1b_2-a_2b_1-a_2b_2-a_1c_1\\
& -a_1c_2-a_2c_1-a_2c_2-b_1c_1-b_1c_2\\
& -b_2c_1-b_2c_2+a_1b_1c_1-a_1b_2c_2\\
& -a_2b_1c_2-a_2b_2c_2+2a_2b_2c_2 \quad\leq 4\\
\end{split}
\end{equation}
This inequality was originally presented in Ref. \cite{J Chen} and said to be violated by all
three-qubit pure entangled states (see Eq. (6) in \cite{J Chen}).

%-----------------------------------------------------------------------------------

\section{Thresholds of Detection Efficiency}
In the actual experiment performed on Bell-type test, there are mainly two kinds of loophole to
overcome, that is, locality loophole and detection loophole \cite{Pearle}. Locality loophole has
been closed in several Bell experiments \cite{Aspect}. However, detection loophole is difficult to
overcome so far \cite{explaination}. In order to perform a loophole-free Bell test, one must make
clear the threshold detection efficiency. Many useful results have been obtained for different Bell
inequalities in two-qubit system \cite{Cabello and Brunner, Garg et.al}. In this section we discuss
the thresholds of detect efficiency for six three-qubit inequalities. Four of them come from Ref.
\cite{Pitowsky} and are labeled from PI-(2) to PI-(5) where the number denotes the equation number
in the reference. The other two inequalities, labeled CI-(2) and CI-(6), are respectively Eq. (2)
and Eq. (6) in Ref. \cite{J Chen}.

We take inequality \eqref{Pito5 prob-form}, i.e., PI-(5), as an illustrative example. As said
before, $P(A_{1})$ is the probability of finding the outcome $\textsf{a}_{1}=+1$, and
$P(A_{1},B_{1})$ is a joint probability of finding the outcomes $\textsf{a}_1=+1$ and
$\textsf{b}_1=+1$. Other probabilities have the similar meaning and all of them are defined in
ideal situation.

Now we assume that the detection efficiencies at site A, B and C are $\eta_{1},\ \eta_{2}$ and
$\eta_{3}$ respectively. Then $P(A_{1})$ should be multiplied by a factor $\eta_{1}$. The joint
probability $P(A_{1},B_{1})$ is replaced by $\eta_{1}\eta_{2}P(A_{1},B_{1})$, and
$P(A_{1},B_{1},C_{1})$ by $\eta_{1}\eta_{2}\eta_{3}P(A_{1},B_{1},C_{1})$. Similarly for other
probabilities. Thus we get the modified form of PI-(5), which is utilized to deal with non-ideal
cases and should be satisfied by LHV model.

In quantum-mechanical formalism, the probability $P(A_{i})$  is replaced by
$\langle\Psi|\mathcal{P}_{a_{1}}^{+}\otimes I\otimes I|\Psi\rangle
=\langle\Psi|\frac{I+A_{1}}{2}\otimes I\otimes I|\Psi\rangle$ for some three-qubit state
$|\Psi\rangle$, and the joint probability $P(A_{1},B_{1})$ by
$\langle\Psi|\mathcal{P}_{a_{1}}^{+}\otimes \mathcal{P}_{b_{1}}^{+}\otimes I|\Psi\rangle
=\langle\Psi|\frac{I+A_{1}}{2}\otimes \frac{I+B_{1}}{2}\otimes I|\Psi\rangle$. Similarly for other
probabilities. Thus we obtain the quantum mechanical expression of the modified PI-(5), that is,
\begin{equation} \label{expectation of J}
\langle\Psi|\mathcal{J}|\Psi\rangle\leq 0
\end{equation}
where $\mathcal{J}$ contains such operators as
$\eta_{1}\mathcal{P}_{a_{1}}^{+},\
\eta_{1}\eta_{2}\mathcal{P}_{a_{1}}^{+}\mathcal{P}_{b_{1}}^{+}$, etc. Once we
find $\langle\Psi|\mathcal{J}|\Psi\rangle >0$, a loophole-free experiment can
be performed. In other words, we hope to find that the maximal eigenvalue of
the operator $\mathcal{J}$ is positive. Before we proceed to numerical
calculation, it is necessary to lower the number of the parameters in
$\mathcal{J}$. Since we are to find the eigenvalues of $\mathcal{J}$, local
observables can be set to be in $xy$-plane. Without loss of generality, we let
\begin{align*}
& A_{1}=B_{1}=C_{1}=\sigma_{x}, \\
& A_{2}=\sigma_{x}\cos\theta_{A}+\sigma_{y}\sin\theta_{A}, \\
& B_{2}=\sigma_{x}\cos\theta_{B}+\sigma_{y}\sin\theta_{B}, \\
& C_{2}=\sigma_{x}\cos\theta_{C}+\sigma_{y}\sin\theta_{C}.
\end{align*}
Thus the $8\times8$ matrix $\mathcal{J}$ contains six parameters: $\eta_{1},\
\eta_{2},\ \eta_{3},\ \theta_{A},\ \theta_{B},\ \theta_{C}$. Then under the
condition that the maximal eigenvalue of $\mathcal{J}$ is positive, we find the
threshold of detection efficiency numerically by means of Mathematica and
MatLab. In the following we discuss two cases and give corresponding results.
\paragraph{Symmetric system} --- In this case, all three detection
efficiencies are the same, i.e., $\eta_{1}=\eta_{2}=\eta_{3}=\eta$. we obtain
the threshold efficiency $\eta^{th}=66.8\%$ on which detection loophole can be
closed.
\paragraph{Asymmetric system} --- In a asymmetric system, three articles
may be detected with different probabilities. We obtain the following results.
\begin{itemize}
\item[(i).] When one of three detectors is perfect, say, $\eta_{1}=1$ and the other two are imperfect
but have the same efficiency, that is, $\eta_{1}=1,\ \eta_{2}= \eta_{3}=\eta$, the threshold
efficiency is found to be $\eta^{th}=50\%$.
\item[(ii).] Two detectors are perfect, e.g., $\eta_{1}=\eta_{2}=1$, and the third is not. In this
case, no matter how the efficiency of the third detector is low, the inequality can be always
violated by appropriate choice of detection orientation.
\item[(iii).] Assuming that $\eta_{1}=1$, just as (i), we consider such a problem: With $\eta_{3}$
arriving at its lower bound, how about the value taken by $\eta_{2}$? Calculation results show that
the lower bound of $\eta_{3}$ can be zero while $\eta_{2}$ must be larger than $50\%$. Obviously
(ii) is the specific case of this result.
\end{itemize}

Inequalities PI-(2) and PI-(4) have similar behavior, whereas inequalities PI-(3), CI-(2) and
CI-(6) do not. For example, for CI-(2) the lower bound of $\eta_{3}$ is $81.9\%$ and can not be
zero, and moreover when $\eta_{3}=81.9\%$ the other two detector must be perfect. All results are
listed in Table \ref{table 1}. Obviously inequality PI-(5) is the most optimal one among them in
the sense that this inequality can endure very low detection efficiency.
\begin{center}
\setlength{\tabcolsep}{0.3cm}
\renewcommand{\arraystretch}{1}
\begin{threeparttable}
\caption{Threshold Detection Efficiency} \label{table 1}%
\begin{tabular}{|c|c|c|c|c|}
\hline \hline &symmetric&\multicolumn{3}{|c|}{asymmetric}\tabularnewline\cline{2-5}&$\eta$&
                                                            $\eta_{1}$&$\eta_{2}$&$\eta_{3}$\\
\hline &&\footnotesize{$1$}&\footnotesize{$50\%$}&\footnotesize{$50\%$}
         \tabularnewline\cline{3-5}\raisebox{1ex}[0pt]{PI-(5)}&\raisebox{1ex}
         [0pt]{\footnotesize{$66.8\%$}}&\footnotesize{$1$}&\footnotesize{$1$}&\footnotesize{$0$}\\
\hline &&\footnotesize{$1$}&\footnotesize{$60\%$}&\footnotesize{$60\%$}
         \tabularnewline\cline{3-5}\raisebox{1ex}[0pt]{PI-(4)}&\raisebox{1ex}
         [0pt]{\footnotesize{$71.5\%$}}&\footnotesize{$1$}&\footnotesize{$1$}&\footnotesize{$0$}\\
\hline &&\footnotesize{$1$}&\footnotesize{$77.2\%$}&\footnotesize{$77.2\%$}
         \tabularnewline\cline{3-5}\raisebox{1ex}[0pt]{PI-(3)}&\raisebox{1ex}
         [0pt]{\footnotesize{$87.6\%$}}&\footnotesize{$1$}&\footnotesize{$1$}&\footnotesize{$34\%$}\\
\hline &&\footnotesize{$1$}&\footnotesize{$66.8\%$}&\footnotesize{$66.8\%$}
         \tabularnewline\cline{3-5}\raisebox{1ex}[0pt]{PI-(2)}&\raisebox{1ex}
         [0pt]{\footnotesize{$87\%$}}&\footnotesize{$1$}&\footnotesize{$1$}&\footnotesize{$0$}\\
\hline &&\footnotesize{$1$}&\footnotesize{$87.4\%$}&\footnotesize{$87.4\%$}
         \tabularnewline\cline{3-5}\raisebox{1ex}[0pt]{CI-(2)}&\raisebox{1ex}
         [0pt]{\footnotesize{$91.5\%$}}&\footnotesize{$1$}&\footnotesize{$1$}&\footnotesize{$76.4\%$}\\
\hline &&\footnotesize{$1$}&\footnotesize{$90.5\%$}&\footnotesize{$90.5\%$}
         \tabularnewline\cline{3-5}\raisebox{1ex}[0pt]{CI-(6)}&\raisebox{1ex}
         [0pt]{\footnotesize{$93.6\%$}}&\footnotesize{$1$}&\footnotesize{$1$}&\footnotesize{$81.9\%$}\\
\hline\hline
\end{tabular}
\end{threeparttable}
\end{center}

\section{Conclusion}
In conclusion, we propose a systematic approach of constructing non-homogeneous Bell-type
inequalities for two- and three-qubit system. In the two-qubit case, non-homogeneous inequality is
attained by subtracting positive projector-like terms from CHSH polynomial. We find that when the
subtracted terms are sufficiently ``large'' the maximal quantum mechanical violation asymptotically
tends to be a constant.

Three-qubit non-homogeneous inequalities are attained by direct generalization of two-qubit ones.
Most of significant three-qubit inequalities presented in literature are recovered in our
framework. The method presented in this paper can be generalized to construct Bell-type
inequalities for multipartite system. The benefit of non-homogeneous inequalities lies in
considering not only full correlations but also partial ones, which are needed in discussing the
non-locality of multipartite system. We conjecture that our method may be used to categorize
Bell-inequalities with various and complicated forms.

Additionally, we analyze numerically the detection efficiency thresholds of the previously
mentioned three-qubit Bell-type inequalities when they are employed to display non-locality in
quantum states. Under different situations, that is, three detectors may have the same or distinct
efficiency, we obtain the thresholds of detection efficiency respectively. For some inequalities,
we find that the efficiency of one detector can be arbitrarily low as long as the other detectors
satisfy certain conditions. Numerical results will help us find the most optimal one which can
endure very low detection efficiency.

\section*{ACKNOWLEDGEMENT}
C.R. is grateful to X. L. Chen for his help. Financial support comes from National Natural Science
Foundation of China, the CAS, Ministry of Education of PRC, and the National Fundamental Research
Program. It is also supported by Marie Curie Action program of the European Union.

%========================================================================================

\end{document}